# Three lines of defense against risks from AI

*Jonas Schuett*[*]

Organizations that develop and deploy artificial intelligence (AI) systems need to manage the associated risks—for economic, legal, and ethical reasons. However, it is not always clear who is responsible for AI risk management. The Three Lines of Defense (3LoD) model, which is considered best practice in many industries, might offer a solution. It is a risk management framework that helps organizations to assign and coordinate risk management roles and responsibilities. In this article, I suggest ways in which AI companies could implement the model. I also discuss how the model could help reduce risks from AI: it could identify and close gaps in risk coverage, increase the effectiveness of risk management practices, and enable the board of directors to oversee management more effectively. The article is intended to inform decision-makers at leading AI companies, regulators, and standard-setting bodies.

## 1 Introduction

Organizations that develop and deploy artificial intelligence (AI) systems need to manage the associated risks—for economic reasons, because accidents and cases of misuse can threaten business performance (Cheatham, Javanmardian, & Hamid Samandari, 2019), for legal reasons, because upcoming AI regulation might require them to implement a risk management system (Schuett, 2022), and for ethical reasons, because AI could have large and long-lasting impacts on society (Clarke & Whitlestone, 2022).

However, it is not always clear who is responsible for AI risk management: The researchers and engineers? The legal and compliance department? The governance team? The Three Lines of Defense (3LoD) model might offer a solution. It is a risk management framework intended to improve an organization's risk governance by assigning and coordinating risk management roles and responsibilities (Institute of Internal Auditors [IIA], 2013, 2020a). It is considered best practice in many industries, such as finance and aviation. In this article, I apply the 3LoD model to an AI context.

To date, there has not been much academic work on the intersection of AI and the 3LoD model. Nunn (2020) suggests using the model to reduce discrimination risks from AI, but the relevant passage is very short. There is also some literature on how companies could use AI to support the three lines (Tammenga, 2020; Sekar, 2022), but I am mainly interested in how to govern AI companies, not how to use AI to govern non-AI companies. It has also been proposed that governments could use the 3LoD model to manage extreme risks from AI (Ord, 2021), but here I focus on the challenges of companies, not government.

---

[*] Research Fellow, Centre for the Governance of AI, Oxford, UK; Research Affiliate, Legal Priorities Project, Cambridge, MA, USA; PhD Candidate, Faculty of Law, Goethe University Frankfurt, Germany; jonas.schuett@governance.ai.

While academic scholarship on this topic may be limited, there is some relevant work from practitioners. Most notably, there is a blog post by PwC that seeks to answer questions similar to this article (Rao & Golbin, 2021). But since they only dedicate a short section to the 3LoD model, their proposal only scratches the surface. The IIA has also published a three-part series, in which they propose an AI auditing framework (IIA, 2017a, 2017c, 2018). Although their proposal contains a reference to the 3LoD model, it does not play a key role. Finally, the 3LoD model is mentioned in a playbook that the National Institute of Standards and Technology (NIST) published alongside the second draft of its AI Risk Management Framework (NIST, 2022a). However, the playbook only suggests implementing the 3LoD model (or a related mechanism), it does not specify how to do so.

Taken together, there are at least two gaps in the current literature. The first one is practical: there does not seem to be a concrete proposal for how organizations that develop and deploy AI systems could implement the 3LoD model. The few proposals that exist are not detailed enough to provide meaningful guidance. The second one is normative: there does not seem to be a thorough discussion about whether implementing the model is even desirable. Given that the model has been criticized and there is not much empirical evidence for its effectiveness, the answer to this question is not obvious. In light of this, the article seeks to answer two research questions: (1) *How could organizations that develop and deploy AI systems implement the 3LoD model?* (2) *To what extent would implementing the 3LoD model help reduce risks from AI?*

The article has three areas of focus. First, it focuses on organizations that develop and deploy state-of-the-art AI systems, in particular medium-sized research labs (e.g. DeepMind and OpenAI) and big tech companies (e.g. Google and Microsoft), though the boundaries between the two categories are blurry (e.g. DeepMind is a subsidiary of Alphabet and OpenAI has a strategic partnership with Microsoft). In the following, I use the term "AI companies" to refer to all of them. I do not cover other types of companies (e.g. hardware companies) or academic institutions, but they might also benefit from my analysis. Second, the article focuses on the organizational dimension of AI risk management. It is not about how AI companies should identify, assess, and respond to risks from AI. Instead, it is about how they should assign and coordinate risk management roles and responsibilities. Third, the article focuses on the model's ability to prevent individual, collective, or societal harm (Smuha, 2021). I am less interested in risks to companies themselves (e.g. litigation or reputation risks), though occasionally private and public interests are aligned (e.g. one way to reduce litigation risks is to prevent accidents).

The remainder of this article proceeds as follows. Section 2 gives an overview of the model's basic structure, history, criticisms, and evidence base. Section 3 suggests ways in which AI companies could implement the model. Section 4 discusses how the model could help reduce risks from AI. Section 5 concludes and suggests questions for further research.

## 2 The 3LoD model

In this section, I give an overview of the basic structure (Section 2.1) and history of the 3LoD model (Section 2.2). I also engage with some of the main criticisms, briefly discuss alternative models (Section 2.3), and review the empirical evidence for its effectiveness (Section 2.4).

### 2.1 Basic structure

There are different versions of the 3LoD model. Most practitioners and scholars are familiar with the version published by the IIA (2013). After a review process, they published an updated version (IIA, 2020a), which increasingly replaces the original version. This article will mainly use the updated version, as illustrated in Figure 1. The updated model has three types of elements: actors, roles, and relationships.

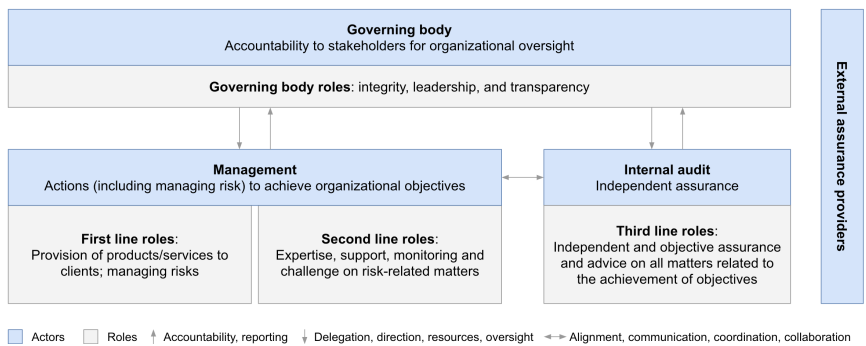

*Figure 1:* The 3LoD model as described by the IIA (2020a)

The model distinguishes between four actors, represented as blue boxes: the governing body, which is accountable to stakeholders for organizational oversight; management, which takes actions to achieve the organization's objectives; internal audit, which provides independent assurance to the governing body, as do external assurance providers.

The model further distinguishes between four roles, represented as gray boxes. The role of the governing body is to demonstrate integrity, leadership, and transparency. In addition to that, the model contains three roles which it calls "lines of defense". The first line provides products and services to clients, and manages the associated risks. The second line assists the first line with regards to risk management. It provides complementary expertise and support, but also monitors and challenges risk management practices. The third line provides independent and objective assurance and advice on all matters related to the achievement of risk objectives. The first two lines are part of management, while the third line is synonymous with internal audit.

Finally, there are three types of relationships between different actors, represented as arrows. There are top-down relationships: the governing body delegates responsibility to management and oversees internal audit. Inversely, there are bottom-up relationships: management and internal audit are accountable and report to the governing body. And lastly, there is a horizontal relationship between actors whose work must be aligned, namely between management and internal audit.

*2.2 Brief history*

The model's origins are opaque. There are theories suggesting military, sporting, or quality control origins (Davies & Zhivitskaya, 2018). It was presumably developed in the late 1990s or early 2000s. In 1999, the Basel Committee on Banking Supervision (BCBS) suggested a similar approach to risk oversight (BCBS, 1999), but the first explicit mention of the model was probably in a report by the UK Financial Services Authority (2003) or a paper by Roman Kräussl (2003).

After the financial crisis of 2007-2008, which was partly caused by widespread risk management failures (Boatright, 2016), the model's popularity skyrocketed. In response to the crisis, regulators and supervisory authorities paid increasing attention to the chief risk officer (CRO) and the risk committee of the board (Walker, 2009; Davies & Zhivitskaya, 2018), and started recommending the 3LoD model (BCBS, 2012; European Banking Authority, 2021). Most academic work on the model was also done after the crisis (e.g. Davies & Zhivitskaya, 2018; Bantleon et al., 2021) and many risk management professionals only heard about the model in its aftermath (Zhivitskaya, 2015).

Today, most listed companies have implemented the 3LoD model. In a 2015 survey of internal audit professionals in 166 countries (n=14,518), the majority of respondents (75%) reported that their organization follows the 3LoD model as articulated by the IIA (Huibers, 2015).[1] Another survey, conducted in 2021 among chief audit executives (CAEs) in Austria, Germany, and Switzerland (n=415), supports their findings (Bantleon et al., 2021). The majority of respondents (88%) reported that they had implemented the model, with particularly high adoption rates among financial institutions (96%).

---

[1] Note that respondents who said they were not familiar with the model were excluded.

In contrast, big tech companies do not seem to have implemented the 3LoD model. It is not mentioned in any of their filings to the US Securities and Exchange Commission (SEC) or other publications. The model is also not explicitly mentioned in the corporate governance requirements by Nasdaq (2022), where all big tech companies are listed. It is worth noting, however, that the risk oversight practices at big tech companies do have some similarities with the 3LoD model. For example, they all seem to have an internal audit function (e.g. Microsoft, 2022; Alphabet, 2022). Based on public information, medium-sized AI research labs do not seem to have implemented the model either.

*2.3 Criticisms and alternative models*

Despite the model's popularity in many industries, it has also been criticized (Arndorfer & Minto, 2015; Zhivitskaya, 2015; Davies & Zhivitskaya, 2018; Hoefer, Cooke, & Curry, 2020; Vousinas, 2021). Arndorfer and Minto (2015) identify four weaknesses and past failures of the 3LoD model. First, they argue, the incentives for risk-takers in the first line are often misaligned. When facing a tradeoff between generating profits and reducing risks, they have historically been incentivized to prioritize the former. Second, there is often a lack of organizational independence for second line functions. They are too close to profit-seekers, which can lead to the adoption of more risk-taking attitudes. Third, second line functions often lack the necessary skills and expertise to challenge practices and controls in the first line. And fourth, the effectiveness of internal audit depends on the knowledge, skills, and experience of individuals, which might be inadequate. Another common criticism is that the model provides a false sense of security. Put simply, "when there are several people in charge—no one really is" (Davies & Zhivitskaya, 2018). Another criticism is that the model is too bureaucratic and costly. Additional layers of oversight might reduce risk, but they come at the cost of efficiency (Zhivitskaya, 2015). A final criticism is that the model depends on information flow between the lines, but there are many barriers to this. For example, the second line might not recognize that they only see what the first line chooses to show them (Zhivitskaya, 2015). While these criticisms identify relevant shortcomings and should be taken seriously, they do not put into question the model as a whole. Moreover, the 3LoD model has been improved over the years. Today, the focus is on increasing the model's effectiveness and responding to criticisms (Davies & Zhivitskaya, 2018).

In view of these criticisms, several alternative models have been suggested. For example, Arndorfer and Minto (2015) proposed the Four Lines of Defense (4LoD) model to better meet the needs of financial institutions. The fourth line consists of supervisory authorities and external audit, who are supposed to work closely with internal audit. Another example is the Five Lines of Assurance (5LoA) model, which was gradually developed by several scholars and organizations (Leech & Hanlon, 2016). However, the proposed changes do not necessarily improve the model. It has been argued that adding more lines would over-complicate the model, and that firms and regulators currently do not want structural changes (Davies & Zhivitskaya, 2018). It is also worth noting that the alternative models are far less popular than the original model. Compared to these alternative models, the 3LoD model remains "the most carefully articulated risk management system that has so far been developed" (Davies & Zhivitskaya, 2018). But what empirical evidence do we have for its effectiveness?

*2.4 Empirical evidence*

By "effectiveness", I mean the degree to which the model helps organizations to achieve their objectives. For the purpose of this article, I am mostly interested in the achievement of risk objectives. This may include: (1) reducing relevant risks to an acceptable level (e.g. risks of harm to individuals, groups, and society), (2) ensuring that management and the board of directors are aware of the nature and scale of key risks, which allows them to define the organization's risk appetite (COSO, 2017), and (3) compliance with relevant risk regulations (Schuett, 2022). I am less interested in other objectives (e.g. improving financial performance), though there might be overlaps (e.g. reducing the risk of harm to individuals might also reduce the risk of financial losses from litigation cases). For an overview of different ways to measure the effectiveness of internal audit, see Rupšys and Boguslauskas (2007), Savčuk (2007), and Boța-Avram and Palfi (2009).

There do not seem to be any (high-quality) studies on the effectiveness of the 3LoD model in the above-mentioned sense.[2] There only seems to be evidence for the effectiveness of internal audit (Lenz & Hahn, 2015; Eulerich & Eulerich, 2020). For example, a survey of CAEs at multinational companies in Germany (n=37) compared audited and non-audited business units within the same company (Carcello et al., 2020). They found that managers of audited units perceive a greater decline in risk compared to managers of non-audited units. Other studies find that internal audit helps to strengthen internal control systems (Lin et al., 2011; Oussii & Taktak, 2018) and has a positive influence on the prevention and identification of fraud (Coram, Ferguson, & Moroney, 2008; Ma'ayan & Carmeli 2016; Drogalas et al., 2017). The fact that the 3LoD model was not able to prevent past scandals and crises seems to provide weak evidence against its effectiveness (though another explanation could be that the model was poorly implemented in these cases), while the model's ongoing popularity seems to provide weak evidence in favor of its effectiveness (though the model's popularity could also be explained by path dependencies). Finally, there is anecdotal evidence in both directions (Zhivitskaya, 2015).

---

[2] There is also not much evidence on the model's effectiveness based on other interpretations of effectiveness. The only exception seems to be a recent study of the 500 largest companies in Denmark, which finds that a higher degree of adherence to first and second line practices is positively associated with financial performance (Andersen, Sax, & Giannozzi, 2022). Besides that, there are only studies on the effects of internal audit (Lenz & Hahn, 2015; Eulerich & Eulerich, 2020; Jiang, Messier, & Wood, 2020), none of which mentions the 3LoD model.

Overall, despite the model's popularity, "its effectiveness [remains] untested" (Davies & Zhivitskaya, 2018) and "not based on any clear evidence" (Power, Ashby, & Palermo, 2013). To be clear, it is not the case that we have robust evidence that the model is ineffective. It is still very plausible that the model can be effective, but there have not been (high-quality) studies providing empirical evidence for its effectiveness in the above-mentioned sense.

This surprising lack of evidence could be explained by the following, rather speculative reasons. First, since it is not feasible to run randomized controlled trials on organizational interventions, it is inherently difficult to collect robust evidence. Second, the model is designed to be flexible and adaptable, which means that there is not a single, standardized way to implement it. This lack of standardization can make it difficult to compare different implementations of the model and to assess their effectiveness.[3] Third, since most practitioners mainly care about financial performance, scholars might be incentivized to focus on that to justify the relevance of their work (though there is not much evidence on that either).

Even if we had more empirical evidence from other industries, its informative value might still be limited. One reason is that findings might not generalize to an AI context. AI companies are structurally different from other companies because they have a special focus on research, and, since AI is a general-purpose technology (Crafts, 2021), risks from AI are broader than risks from other products and services. Another reason is that the biggest driver of the model's ability to reduce risks is likely the concrete way in which it is implemented. Instead of asking "is the 3LoD model effective?", AI companies should ask "how can we implement the model in an effective way?".

## 3 Applying the 3LoD model to an AI context

This section suggests ways in which AI companies could implement the 3LoD model. For each of the three lines, I suggest equivalent roles and responsibilities. First, I describe the content of their responsibilities, then I discuss which team or individual would be responsible, as illustrated in Figure 2.

---

[3] This argument was suggested by OpenAI's language model GPT-3.

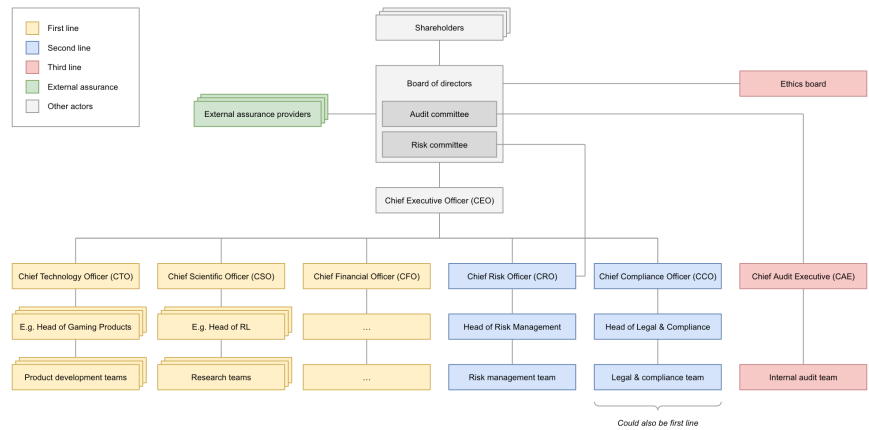

*Figure 2:* Sample org chart of an AI company with equivalent responsibilities for each of the three lines

## 3.1 First line

The first line has two main responsibilities: providing products and services to clients, which corresponds to AI research and product development, and managing the associated risks. Below, I focus on the latter.

The first line is responsible for establishing and maintaining appropriate structures and processes for the management of risk. This involves measures along all steps of the risk management process. For example, to identify risks from AI, the first line could use risk taxonomies (Microsoft, 2020; Weidinger et al., 2021; Raji et al., 2022), incident databases (McGregor, 2021), or scenario planning and wargaming (International Electrotechnical Commission [IEC], 2019; Gyengo & Bruner, 2022). To estimate the likelihood and severity of the identified risks, and to assess potential vulnerabilities, the first line might use Bayesian networks, Monte Carlo simulations, or penetration testing (IEC, 2019; International Organization for Standardization [ISO] & IEC, 2022). To reduce risks, it could fine-tune the model on a curated dataset (Solaiman & Dennison, 2021), introduce a policy for the publication of potentially harmful research (Partnership on AI, 2021; Solaiman et al., 2019), or only grant structured access to models (e.g. via an API) to reduce misuse risks (Shevelane, 2022). The first line could also take a more holistic approach and implement an AI-specific risk management framework (e.g. NIST, 2022b; ISO & IEC, n.d.) or customize a more general enterprise risk management (ERM) framework (e.g. ISO, 2018; Committee of Sponsoring Organizations of the Treadway Commission [COSO], 2017).

The first line is also responsible for ensuring compliance with legal, regulatory, and ethical expectations. Legal obligations might stem from anti-discrimination law (Wachter, Mittelstadt, & Russell, 2021), data protection law (Hamon et al., 2022), or antitrust law (Hua & Belfied, 2021). A notable example of AI regulation is the proposed EU AI Act (European Commission, 2021), which requires providers of high-risk AI systems to implement a risk management system (Schuett, 2022). Ethical expectations might stem from AI ethics principles that organizations have adopted on a voluntary basis (Jobin, Ienca, & Vayena, 2019). To ensure compliance, the first line relies on support from the second line (see below).

Finally, the first line is responsible for informing the governing body about the outcomes of the above-mentioned measures, the degree to which risk objectives are met, and the overall level of risk. This should take the form of a continuous dialogue, including reporting about expected and actual outcomes. Reporting will typically include heat maps and risk registers (IEC, 2019), but it could also involve information about specific models, in the form of model cards (Mitchell et al., 2018), data sheets (Gebru et al., 2018), and system cards (Green et al., 2021). Note that there should also be a reporting line from the CRO to the chief executive officer (CEO) and the risk committee of the board (see below).

Responsible are operational managers, often in a cascading responsibility structure. At big tech companies, the lowest level of responsibility would lie with those managers who are in charge of the development of individual AI products. If there is no stand-alone AI product and AI systems make up only part of a product (e.g. WaveNet as a part of Google Assistant), then the lowest level of responsibility would lie with those managers who lead the development of the AI part of the product (e.g. the research lead for WaveNet). At medium-sized research labs, the lowest level of responsibility for risk management would lie with research leads, i.e. senior researchers who are in charge of individual research projects.

There will usually be one or more intermediate levels of responsibility. This might include a number of mid-level managers responsible for broader product areas (e.g. gaming) or research areas (e.g. reinforcement learning), though the details depend on the particular organizational structures. The ultimate responsibility for AI risk management lies with those C-suite executives who are responsible for product development (e.g. the chief technology officer [CTO]) or research (e.g. the chief scientific officer [CSO]). While it is possible to split responsibilities between two or more executives, this is often not advisable, mainly because it can dilute responsibilities.

*3.2 Second line*

The second line is responsible for assisting the first line with regards to risk management. It provides complementary expertise and support, but also monitors and challenges risk management practices.

Some risk management activities require special expertise that the first line does not have. This might include legal expertise (e.g. how to comply with the risk management requirements set out in the proposed EU AI Act [Schuett, 2022]), technical expertise (e.g. how to develop more truthful language models [Evans et al., 2021]), or ethical expertise (e.g. how to define normative thresholds for fairness [Kleinberg, Mullainathan, & Raghavan, 2016]). It might also include risk-specific expertise (e.g. what risks language models pose [Weidinger et al., 2021]) or risk management-specific expertise (e.g. best practices for red teaming safety filters [Rando et al., 2022]). The second line could support the first line by drafting policies, processes, and procedures, as well as frameworks, templates, and taxonomies. It might also advise on specific issues (e.g. how to customize a risk management framework to better meet the specific needs of the company), provide general guidance (e.g. how to ensure compliance with safety-related policies among researchers and engineers), or offer trainings (e.g. how to process training data in a GDPR compliant way).

The second line is also responsible for monitoring and challenging the adequacy and effectiveness of risk management practices. Risk management practices are ineffective if risk objectives are not met (e.g. the company fails to comply with relevant laws and regulations, or it is unable to reduce risks to an acceptable level). They are inadequate if the same results could have been achieved with fewer resources. The second line will typically use a number of key performance indicators (KPIs) to evaluate various dimensions of the adequacy and effectiveness of risk management (e.g. number of identified risks, number of incidents, or percentage of personnel trained on specific matters).

Second line responsibilities are split across multiple teams. This typically includes the risk management team as well as the legal and compliance team. Although most big tech companies already have a risk management team, these teams are mostly concerned with business risks (e.g. litigation or reputation risk). Risks from AI, especially societal risks, are usually not a major concern (Smuha, 2021). If big tech companies want to change this, they could expand the responsibilities of existing teams. Setting up a new AI-specific risk management team seems less desirable, as it could lead to a diffusion of responsibilities. There would likely be a cascading responsibility structure where the CRO acts as the single point of accountability for the risk management process. Medium-sized research labs usually do not have a dedicated risk management team. They could either set up a new team or task one or more people in other teams with risk management-related support functions.

All AI companies beyond the early startup phase have a legal and compliance team. The team lead, and ultimately the chief compliance officer (CCO) or chief legal officer (CLO), would be responsible for risk-related legal and compliance support. It is worth noting that the legal and compliance team can also be part of the first line if they are actually responsible for ensuring compliance. They are part of the second line if they do not have any decision power and only support the first line (e.g. by writing legal opinions). The legal and compliance team can also seek support from external law firms.

Many organizations that develop and deploy AI systems have other teams that could take on second line responsibilities. This might include technical safety, ethics, policy, or governance teams. However, in practice, these teams rarely consider themselves as being responsible for risk management. This needs to be taken into account when implementing the 3LoD model (e.g. by running workshops to sensitize them to their widened responsibility). In general, AI companies should arguably avoid assigning second line responsibilities to them.

*3.3 Third line*

The third line is responsible for providing independent assurance. It assesses the work of the first two lines and reports any shortcomings to the governing body.

While the second line already monitors and challenges the adequacy and effectiveness of the risk management practices, the third line independently assesses their work—they supervise the supervisors, so to speak. They could do this by conducting an internal audit (Raji et al, 2020) or commissioning an external audit (Buolamwini & Gebru, 2018; Mökander & Floridi, 2022). Such audits could have different purposes and scopes (Mökander et al., 2022). They could evaluate compliance with laws, standards, or ethics principles ("compliance audit") or seek to identify new risks in a more open-ended fashion ("risk audit"). They could also assess the model itself, including the dataset it was trained on ("model audit"), the model's impact ("impact audit"), or the company's governance ("governance audit"). Similarly, the third line could engage a red team before or after a model is deployed to assess if the first two lines were able to identify all relevant risks (Ganguli et al., 2022; Perez et al., 2022). For example, before OpenAI released DALL·E 2, they asked a group of external experts to identify ways in which the model can be misused (Mishkin et al., 2022). In addition to that, the third line should also review key policies and processes to find flaws and vulnerabilities (e.g. ways in which a policy that requires researchers to assess the societal impact of a model can be circumvented [Ashurst et al., 2022]). Note that this should also include a meta-assessment of the company's implementation of the 3LoD model itself.

The third line also supports the governing body, typically the board of directors, by providing independent and objective information about the company's risk management practices (IIA, 2020b). Their main audience is usually the audit committee, which is mainly composed of non-executive directors. But since non-executive directors only work part-time and heavily depend on the information provided to them by the executives, they need an independent ally in the company to effectively oversee the executives (Davies & Zhivitskaya, 2018). The third line serves this function by maintaining a high degree of independence from management and reporting directly to the governing body following best practices. It is often described as their "eyes and ears" (IIA, 2020a).

The third line has a well-defined organizational home: internal audit. Note that, in this context, internal audit refers to a specific organizational unit. It does not merely mean an audit that is done internally (Raji et al, 2020). Instead, it means "those individuals operating independently from management to provide assurance and insight on the adequacy and effectiveness of governance and the management of risk (including internal control)" (IIA, 2020a).

Typically, companies have a dedicated internal audit team, led by the CAE or Head of Internal Audit. Most big tech companies have such a team, but similar to the risk management team, they often neglect the societal risks from AI. Instead of creating a separate AI-specific internal audit team, they should create a sub-team within their existing internal audit team, or simply task one or more team members to focus on AI-specific risk management activities. Medium-sized research labs usually do not have an internal audit team. They would have to create a new team or task at least one person with third line responsibilities. In short, big tech companies need to "bring AI to internal audit", while research labs need to "bring internal audit to AI". It is worth noting that, although there are promising developments (IIA, 2017a, 2017c), the profession of AI-specific internal auditors is still in its infancy.

Some AI companies have an ethics board (e.g. Microsoft's Aether Committee and Facebook's Oversight Board) which could also take on third line responsibilities, typically in addition to internal audit. It would have to be organizationally independent from management, but still be part of the organization (in contrast to external assurance providers). If organizations already have an independent ethics board (e.g. consisting of representatives from academia and civil society), they could form a working group that takes on third line responsibilities.

## 4 How the 3LoD model could help reduce risks from AI

While there are many reasons why AI companies may want to implement the 3LoD model, this section focuses on three arguments about the model's ability to prevent individual, collective, and societal harm: the model could help reduce risks from AI by identifying and closing gaps in risk coverage (Section 4.1), increasing the effectiveness of risk management practices (Section 4.2), and enabling the governing body to oversee management more effectively (Section 4.3). I also give an overview of other benefits (Section 4.4). It is worth noting that, in the absence of robust empirical evidence (see above), the following discussion remains theoretical and often relies on abstract plausibility considerations.

*4.1 Identifying and closing gaps in risk coverage*

AI risk management involves different people from different teams with different responsibilities (Baquero et al., 2020). If these responsibilities are not coordinated adequately, gaps in risk coverage can occur (Bantleon et al., 2021). Such gaps may have different causes. For example, it might be the case that no one is responsible for managing a specific risk (e.g. there could be a blind spot for diffuse risks), or it might be unclear who is responsible (e.g. two teams might incorrectly assume that the other team already takes care of a risk). Gaps could also occur if the responsible person is not able to manage the risk effectively (e.g. because they do not have the necessary expertise, information, or time). If a specific risk is not sufficiently covered by the risk management system, it cannot be identified, which might result in an incorrect risk assessment (e.g. the total risk of an unsafe AI system is judged acceptable) and an inadequate risk response (e.g. an unsafe AI system is deployed without sufficient safety precautions).

The 3LoD model could prevent this by identifying and closing gaps in risk coverage. It could do this by offering a systematic way to assign and coordinate risk management-related roles and responsibilities. It ensures that people who are closest to the risk are responsible for risk management (first line) and get the support they need (second line). Another way the 3LoD model can help identify blindspots is through the internal audit function (third line). They are responsible for assessing the adequacy and effectiveness of the entire risk management regime, which includes potential gaps in risk coverage.

One might object that, in practice, gaps in risk coverage are rare, and even if they occur, they only concern minor risks (e.g. because AI companies have found other ways to address the biggest risks). However, the AI Incident Database (McGregor, 2021) contains numerous entries, including several cases classified as "moderate" or "severe", which indicates that incidents are not that uncommon. While these incidents had many different causes, it seems plausible that at least some of them were related to gaps in risk coverage. But since there does not seem to be any public data on this, the issue remains speculative.

Even if one thinks that gaps in risk coverage are a common problem among AI companies, one might question the model's ability to identify and close them. One might suspect that the people involved and their ability and willingness to identify gaps play a much bigger role. While it is certainly true that implementing the model alone is not sufficient, neither is having able and willing personnel. Both are necessary and only together can they be sufficient (though other factors, such as information sharing between different organizational units, might also play a role).

Overall, it seems likely that implementing the 3LoD model would help uncover some gaps in risk coverage that would otherwise remain unnoticed.

*4.2 Increasing the effectiveness of risk management practices*

Some risk management practices are ineffective—they might look good on paper, but do not work in practice. AI companies might fail to identify relevant risks, misjudge their likelihood or severity, or be unable to reduce them to an acceptable level. Ineffective risk management practices can have many different causes, such as reliance on a single measure (e.g. using a single taxonomy to identify a wide range of risks), a failure to anticipate deliberate attempts to circumvent measures (e.g. stealing an unreleased model), a failure to anticipate relevant changes in the risk landscape (e.g. the emergence of systemic risks due to the increasing reliance on so-called "foundation models" [Bommasani et al., 2021]), cognitive biases of risk managers (e.g. the availability bias, i.e. the tendency to "assess the frequency of a class or the probability of an event by the ease with which instances or occurrences can be brought to mind" [Tversky & Kahneman, 1974]), and other human errors (e.g. a person filling out a risk register slips a line), among other things.

The 3LoD model can increase the effectiveness of risk management practices by identifying such shortcomings. As mentioned above, internal auditors assess the effectiveness of risk management practices (e.g. via audits or red teaming exercises) and report any shortcomings to the governing body, which can engage with management to improve these practices.

One might object that most shortcomings only occur in low-stakes situations. In high-stakes situations, existing risk management practices are already more effective. For example, AI companies often conduct extensive risk assessments before deploying state-of-the-art models (Brundage et al., 2022; Kavukcuoglu et al., 2022). While this might be true in obvious cases, there are less obvious cases where practices might not be as effective as intended (e.g. because they are insensitive to human errors or deliberate attempts to circumvent them). I would certainly not want to rely on the counterargument that the effectiveness of risk management practices already scales sufficiently with the stakes at hand.

Some AI companies might further object that they already have the equivalent of an internal audit function, so implementing the 3LoD would only be a marginal improvement. While it might be true that some people at some companies perform some tasks that are similar to what internal auditors do, to the best of my knowledge, assessing the effectiveness of risk management practices is not their main responsibility and they do not follow best practices from the internal audit profession, such as being organizationally independent from management (IIA, 2017b), which can lead to noticeable differences.

Overall, I think this is one of the best arguments for implementing the 3LoD model. Without a serious attempt to identify ineffective risk management practices, I expect at least some shortcomings to remain unnoticed. The degree to which this is true mainly depends on internal audit's ability and willingness to serve this function.

### 4.3 Enabling the governing body to oversee management more effectively

The governing body, typically the board of directors, is responsible for overseeing management. To do this, they need independent and objective information about the company's risk management practices. However, they heavily rely on information provided to them by the executives. To effectively oversee the executives, they need an independent ally in the company.

Internal audit serves this function by maintaining a high degree of independence from management and reporting directly to the audit committee of the board. This can be important because, compared to other actors, the board has significant influence over management. For example, they can replace the CEO (e.g. if they repeatedly prioritize profits over safety), make strategic decisions (e.g. blocking a strategic partnership with the military), and make changes to the company's risk governance (e.g. setting up an ethics board). Note that there is a complementary reporting line from the CRO to the risk committee of the board.

One might object that this function could also be served by other actors. For example, third-party auditors could also provide the board with independent and objective information. While external audits can certainly play an important role, they have several disadvantages compared to internal audits: they might lack important context, companies might not want to share sensitive information with them (e.g. about ongoing research projects), and audits are typically only snapshots in time. AI companies should therefore see external audit as a complement to internal audit, not a substitution. There is a reason why the 3LoD model distinguishes between internal audit and external assurance providers.

One might further point out that in other industries, internal audit is often perceived to intervene too late (Davies & Zhivitskaya, 2018) and to team up with management, instead of monitoring them (Roussy & Rodrigue, 2018). This would indeed be problematic. However, as discussed above, this does not seem to be an inherent property of internal audit. Instead, it seems to be mainly driven by the particular way it is set up and the people involved. Having said that, AI companies should take this concern seriously and take measures to address it.

Overall, I think that implementing the 3LoD model can significantly increase the board's information base. This effect will be more noticeable at medium-sized research labs, as most big tech companies already have an internal audit function, albeit not an AI-specific one (see above).

*4.4 Other benefits*

Implementing the 3LoD model has many benefits other than reducing risks to individuals, groups, or society. Although these other benefits are beyond the scope of this article, it seems warranted to at least give an overview. Below, I briefly discuss four of them.

First, implementing the 3LoD model can avoid unnecessary duplications of risk coverage. Different people in different teams could be doing the same or very similar risk management work. This is often desirable because it can prevent gaps in risk coverage (see above). But if such duplications are not necessary, they can waste resources, such as labor, that could be used more productively elsewhere. AI companies therefore face an effectiveness-efficiency-tradeoff. How this tradeoff ought to be resolved, depends on the particular context. For example, when dealing with catastrophic risks, effectiveness (preventing gaps in risk coverage) seems more important than efficiency (avoiding unnecessary duplications of coverage). In this case, AI companies should strictly err on the side of too much coverage rather than risk gaps in important areas. Overall, this benefit seems to be overstated and less relevant if one is mainly concerned with risk reduction.

Second, AI companies that have implemented the 3LoD model might be perceived as being more responsible. In general, risk management practices at AI companies seem less advanced compared to many other industries (e.g. aviation or banking). By adapting existing best practices from other industries, they would signal that they aim to further professionalize their risk management practices, which could be perceived as being more responsible. This perception might have a number of benefits. For example, it could make it easier to attract and retain talent that cares about ethics and safety. It could also help avoid overly burdensome measures from regulators. It might even be beneficial in litigation cases for the question of whether or not an organization has fulfilled its duty of care. However, it seems questionable whether implementing the 3LoD model affects perception that much, especially compared to other governance measures (e.g. publishing AI ethics principles or setting up an AI ethics board), mainly because most stakeholders, including most employees, do not know the model and cannot assess its relevance. An exception might be regulators and courts who care more about the details of risk management practices. My best guess is that implementing the model will have noticeable effects on the perception of a few stakeholders, while most other stakeholders will not care.

Third, implementing the 3LoD model can make it easier to hire risk management talent. The profession of AI risk management is in its infancy. I assume that AI companies find it challenging to hire people with AI *and* risk management expertise. In most cases, they can either hire AI experts and train them in risk management, or hire risk management experts from other industries and train them in AI. Implementing the 3LoD model could make it easier to hire risk management experts from other industries because they would already be familiar with the model. This might become more important if one assumes that AI companies will want to hire more risk management talent as systems get more capable and are used in more safety-critical situations (e.g. Degrave et al., 2022). However, I do not find this argument very convincing. I doubt that implementing the 3LoD model would make a meaningful difference on relevant hiring decisions (e.g. on a candidate's decision to apply or accept an offer). Since the model is about the organizational dimension of risk management, it does not have significant effects on the day-to-day risk management work. Having said that, there might be smaller benefits (e.g. making the onboarding process easier). My best guess is that the counterfactual impact of 3LoD implementation on hiring is low.

Fourth, implementing the 3LoD model might reduce financing costs. Rating agencies tend to give better ratings to companies that have implemented an ERM framework (because doing so is considered best practice), and companies with better ratings tend to have lower financing costs (because they get better credit conditions) (see Bohnert et al., 2019). There might be an analogous effect with regards to the implementation of the 3LoD model. Lower financing costs are particularly important if one assumes that the costs for developing state-of-the-art AI systems will increase because of increasing demand for compute (Sevilla et al., 2022), for example. In scenarios where commercial pressure is much higher than today, lower financing costs could also be important to continue safety research that does not contribute to product development. That said, I am uncertain to what extent the findings for ERM frameworks generalize to the 3LoD model. My best guess is that implementing the 3LoD would not have meaningful effects on the financing costs of medium-sized research labs today. But I expect this to change as labs become more profitable and increasingly make use of other funding sources (e.g. credits or bonds).

## 5 Conclusion

This article has applied the 3LoD model to an AI context. It has suggested concrete ways in which medium-sized research labs like DeepMind and OpenAI or big tech companies like Google and Microsoft could implement the model to reduce risks from AI. It has argued that implementing the model could prevent individual, collective, or societal harm by identifying and closing gaps in risk coverage, increasing the effectiveness of risk management practices, and enabling the governing body to oversee management more effectively. It concluded that, while there are some limitations and the effects should not be overstated, the model can plausibly contribute to a reduction of risks from AI.

Based on the findings of this article, I suggest the following questions for further research. First, the article has highlighted the importance of internal audit in AI risk management. But since there has not been much AI-specific work on internal audit, it would be valuable to review best practices from other industries (e.g. BCBS, 2012) and discuss the extent to which these practices are applicable to an AI context. Second, my discussion of the model's ability to reduce risks from AI was mostly theoretical and relied on abstract plausibility considerations. I encourage other scholars to assess these claims empirically. An industry case study similar to the one that Mökander and Floridi (2022) conducted for ethics-based auditing could be a first step. Third, although AI companies have not implemented the 3LoD model, they already perform many of the above-mentioned activities. To better target future work, it would be helpful to review existing risk management practices at leading AI companies and conduct a gap analysis. Since public data is scarce, scholars would have to conduct interviews or surveys (e.g. an "AI risk management benchmark survey"), though I expect confidentiality to be a major obstacle. Fourth, the article has focused on the voluntary adoption of the 3LoD model. It would be important to know if existing or future regulations might even require AI companies to implement the model. For example, while Article 9 of the proposed EU AI Act does not mention the 3LoD model, it has been suggested that future harmonized standards or common specifications should include the model (Schuett, 2022). The 3LoD model is also mentioned in the playbook that accompanies the NIST AI Risk Management Framework (NIST, 2022a, 2022b). It is conceivable that this framework will be translated into US law, similar to the NIST Framework for Improving Critical Infrastructure Cybersecurity (NIST, 2018). Finally, the article has investigated the 3LoD in isolation. It has excluded contextual factors, such as the risk culture at AI companies, which might also affect the model's effectiveness. A better understanding of these factors would further improve the information base for decision-makers at AI companies and beyond.

As famously put by George Box (1976), "all models are wrong, but some are useful". In the same spirit, one might say that the 3LoD model is not a silver bullet against the risks from AI, but it can still play an important role. AI companies should see it as one of many governance tools they can use to tackle today's and tomorrow's threats from AI.

## Acknowledgements

I am grateful for valuable comments and feedback from Leonie Koessler, James Ginns, Markus Anderljung, Andre Barbe, Noemie Dreksler, Toby Shevelane, Anne le Roux, Alexis Carlier, Emma Bluemke, Christoph Winter, Renan Araújo, José Jaime Villalobos, Suzanne Van Arsdale, Alfredo Parra, and Nick Hollman. All remaining errors are my own.